\documentclass[%
reprint,
superscriptaddress,
 amsmath,amssymb,
prb,
]{revtex4-2}

\usepackage{graphicx}
\usepackage{dcolumn}
\usepackage{bm}
\usepackage{hyperref}
\usepackage{cleveref}
\hypersetup{
    colorlinks = true,
    allcolors=black,
    citecolor=blue,
    linkcolor=blue
} 
\usepackage{amsmath, amsfonts, dsfont, mathrsfs, float,amssymb}
\usepackage{color}
\usepackage{tikz}
\usepackage{tikz-network}

\newcommand{\barj}[1]{\bar{\jmath}}

\renewcommand{\vec}[1]{\mathbf{#1}}

\begin{document}


\title{An Entanglement-Complexity Generalization of the Geometric Entanglement}

\author{Alexander Nico-Katz}
\affiliation{Department of Physics and Astronomy, University College London, London WC1E 6BT, United Kingdom}
\author{Sougato Bose}
\affiliation{Department of Physics and Astronomy, University College London, London WC1E 6BT, United Kingdom}

\date{\today}
             
\begin{abstract}
    We propose a class of generalizations of the geometric entanglement for pure states by exploiting the matrix product state formalism. This generalization is completely divested from the notion of separability and can be freely tuned as a function of the bond dimension to target states which vary in entanglement complexity. We first demonstrate its value in a toy spin-1 model where, unlike the conventional geometric entanglement, it successfully identifies the AKLT ground state. We then investigate the phase diagram of a Haldane chain with uniaxial and rhombic anisotropies, revealing that the generalized geometric entanglement can successfully detect all its phases and their entanglement complexity. Finally we investigate the disordered spin-$1/2$ Heisenberg model, where we find that differences in generalized geometric entanglements can be used as lucrative signatures of the ergodic-localized entanglement transition.
\end{abstract}
             
\maketitle
\section{Introduction}
\label{sec:intro}
The role of entanglement in many body systems has become one of the most important topics in modern quantum physics. 
The geometric measure of entanglement - defined as the distance of a state from the nearest separable state - and its $k$-separable generalizations wherein the state is instead separable into $k$ subsystems, have been particularly valuable and have seen widespread use in quantum information and in the investigation of bi-partite and multi-partite entanglement structures \cite{Wei2003, Shi2010, SenDe2010, Biswas2014, Singha2019}. However, many systems yield states which have interesting entanglement structures that are not separable into any partitions of the system into subsystems. Examples include the non-separable AKLT state which saturates the geometric entanglement \cite{Orus2008}, or the many body localized eigenstates which are complicated but area-law entangled \cite{Alet2018}. Thus separability alone, even when we extend the definition to separability between arbitrarily partitioned subsystems, is not comprehensive enough to characterize how much entanglement is required to accurately describe a many-body state. Thus, one can naturally ask the question: can a generalization of the geometric entanglement be constructed that goes beyond separability?

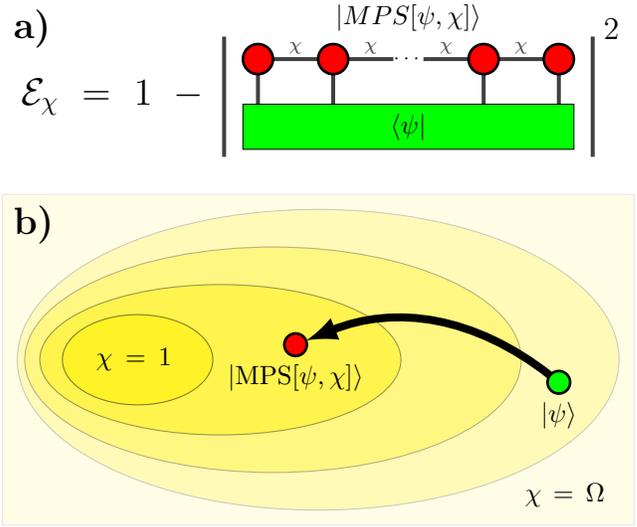
\begin{figure}[ht]
    \centering
    \begin{tikzpicture}
        \Vertex[x=-.8, y=4., color=red, size=.4]{node1}
        \Vertex[x=0.2, y=4., color=red, size=.4]{node2}
        \Vertex[x=1.2, y=4., color=red, size=.4, label=$\cdots$,Pseudo]{nodemid}
        \Vertex[x=2.2, y=4., color=red, size=.4]{node3}
        \Vertex[x=3.2, y=4., color=red, size=.4]{node4}
        \Vertex[x=-.8, y=3.,Pseudo]{node1d}
        \Vertex[x=0.2, y=3.,Pseudo]{node2d}
        \Vertex[x=2.2, y=3.,Pseudo]{node3d}
        \Vertex[x=3.2, y=3.,Pseudo]{node4d}
        \draw[fill=green] (-1., 2.8) rectangle (3.4, 3.4);
        \Edge[label=$\chi$,position=above](node1)(node2)
        \Edge[label=$\chi$,position=above](node2)(nodemid)
        \Edge[label=$\chi$,position=above](nodemid)(node3)
        \Edge[label=$\chi$,position=above](node3)(node4)
        \Edge(node1)(node1d)
        \Edge(node2)(node2d)
        \Edge(node3)(node3d)
        \Edge(node4)(node4d)

        \Vertex[x=3.65,y=4.,Pseudo]{rv}
        \Vertex[x=-1.25,y=4.,Pseudo]{lv}
        \Vertex[x=3.65,y=3.,Pseudo]{rvb}
        \Vertex[x=-1.25,y=3.,Pseudo]{lvb}
        \Edge(rv.north)(rvb.south)
        \Edge(lv.north)(lvb.south)
        \Text[x=3.90,y=4.4,fontsize=\large]{2}
        
        \Text[x=1.2,y=3.1]{$\langle\psi|$}
        \Text[x=1.2,y=4.55]{$|MPS{[\psi,\chi]}\rangle$}
        
        \Text[x=-2.6,y=3.5,fontsize=\Large]{$\mathcal{E}_\chi~=~1~-~$}
        
        \draw[fill=yellow,opacity=.1] (-4.2cm,-2.2cm) rectangle (4.2cm, 2.2cm);
        \draw[fill=yellow,opacity=.2] (0,0) ellipse (4cm and 2cm); 
        \draw[fill=yellow,opacity=.3] (-.6,0) ellipse (3.3cm and 1.5cm); 
        \draw[fill=yellow,opacity=.4] (-1.3,0) ellipse (2.4cm and 1cm); 
        \draw[fill=yellow,opacity=.5] (-2.4,0) ellipse (1.0cm and .6cm);
        
        \Vertex[x=-.3,y=0.2,size=.3,color=red,label=|\text{MPS}{[\psi,\chi]}\rangle,position=below,Math,fontsize=\normalsize]{state}
        \Vertex[x=3.2,y=-.3,size=.3,color=green,label=|\psi\rangle,position=below,Math,fontsize=\normalsize]{mps}
        \Text[x=-2.2, y=0, width = 1.5cm]{$\chi = 1$}
        \Text[x=3.5, y=-1.8, width = 1.5cm]{$\chi = \Omega$}
        \Text[x=-3.3, y=1.8, width = 1.5cm,fontsize=\Large]{\textbf{b)}}
        \Text[x=-3.3, y=4.4, width = 1.5cm,fontsize=\Large]{\textbf{a)}}
        \Edge[bend=-30,Direct,color=black,opacity=1,lw=3,NotInBG](mps)(state) 

    \end{tikzpicture}
    \caption{Schematics showing \textbf{a)} the diagrammatic equation for our generalization of the geometric entanglement (an overview of this diagrammatic notation is given in \cref{sec:app-diagram}) and \textbf{b)} a Hilbert space which has been organized into nested manifolds of states with perfect representations as MPS of bond dimension $\chi$. The $\chi=1$ manifold is a manifold of product states, and the full Hilbert space is attained as $\chi \to \Omega$ where $\Omega$ is the total dimension of the space. The compression procedure of a state $|\psi\rangle$ into its MPS representation $|\text{MPS}[\psi,\chi]\rangle$ is given by the black arrow.}
    \label{fig:schematic}
\end{figure}

We address this question by proposing a new geometric measure based on the entanglement-complexity perspective of the matrix product state (MPS) formalism. The central object of this formalism, the MPS itself, is an alternative representation of a generic pure state. This representation becomes advantageous when the amount of entanglement in a state is limited in some way; if this is the case then much of the exponentially large Hilbert space is irrelevant and can be discarded \cite{Vidal2003}. This truncation of Hilbert space is controlled by the bond dimension $\chi$ which quantifies how much information the MPS representation of a state can contain. States with a low amount of entanglement permit exact (or close to exact) representations as MPS of low bond dimension, whilst states with a large amount of complicated entanglement structures require an MPS of large bond dimension that approaches the dimension of the total original Hilbert space \cite{Schollwock2011, Orus2014}. We thus generalize the geometric entanglement without appealing to separability by reversing this approach: the entanglement of a given state can be quantified by how much a low-$\chi$ MPS \textit{fails} to represent it.

We begin by reviewing the relevant aspects of the MPS formalism in \cref{sec:mps}, namely the decomposition procedure of a state into its MPS representation and the origin and interpretation of the bond dimension $\chi$. In \cref{sec:generalizing} we discuss the geometric entanglement, its extant generalization in terms of $k$-separability, and introduce our MPS-theoretic generalization. The rest of the paper consists of an examination of different systems in which our generalization exhibits a clear advantage over the conventional geometric entanglement; establishing it as a vital tool in exploratory analysis of quantum systems. In \cref{sec:aklt} we examine the spin-1 Haldane model across the AKLT point: a setting in which separability is not a useful signature of entanglement, and is thus the ideal setting in which to introduce and justify our generalization. In \cref{sec:haldane} we perform an exploratory investigation of the phase diagram of a spin-1 Haldane chain with uniaxial and rhombic anisotropies. This investigation reveals that using different values of the bond dimension $\chi$ in our generalization yields different phase diagrams, with each value of $\chi$ revealing new phases or features. This establishes a class of generalizations which together form a set of highly tunable exploratory tools. Finally in \cref{sec:mbl} we investigate the many body localization transition in a spin-1/2 Heisenberg chain, establishing the value of our generalization away from ground state transitions. Together, these investigations demonstrate a range of contexts in which our generalization of the geometric entanglement exhibits striking advantages; offering an approach by which systems with limited but non-separable entanglement structures can be investigated using geometric measures.

\section{The MPS Representation}
\label{sec:mps}

The central object of the MPS formalism is the MPS itself: a representation of an arbitrary pure state $|\psi\rangle$ as a product of local tensors. This representation is formed by repeated reshaping and decomposition of the original state until it has been factorized into the MPS form, a process we review here. Starting from the generic pure state
\begin{equation}\label{eq:initial}
    |\psi\rangle = \sum_{\{j\}}^d c_{j_1, j_2, \cdots, j_N} |j_1, j_2, \cdots, j_n\rangle,
\end{equation}
where the $j$ indices are `physical' indices which account for physical degrees of freedom, we combine the indices $j_2, j_3 \cdots, j_N$, reshape the tensor, and perform a singular value (Schmidt) decomposition across the physical indices $j_1$ and $(j_2, j_3, \cdots, j_N)$:
\begin{equation}
    c_{j_1, (j_2, \cdots, j_N) } = \sum_{s_{2}} {U_{j_1, s_2} S_{s_2, s_2} V^\dagger_{s_2, (j_2,j_3, \cdots, j_N) }}.
\end{equation}
The matrix $U$ is left-unitary, $V^\dagger$ is right-unitary, and $S$ is a diagonal matrix of the descending singular values across the bi-partition, of which some may be degenerate or exactly zero.

These singular values determine the quality of the decomposition: low singular values contribute less to the decomposition and can be discarded without a significant decrease in the fidelity of our MPS representation. The bond dimension $\chi$ is the positive integer number of these singular values that we choose to keep and quantifies the amount of information retained by our MPS. In general, the more singular values we discard at every partition, the more compressed and less exact our MPS representation becomes. It is here that the concept of entanglement complexity becomes important: since the number of singular values across a bi-partition is an entanglement measure in its own right, states with less entanglement have more singular values equal or close to zero that can be readily discarded, and have correspondingly good low-$\chi$ MPS representations \cite{Eisert2001, Vidal2003}. The more entanglement there is in a system, and the more complicated its entanglement structure is, the higher the bond dimension required to achieve a good MPS representation \cite{Orus2014}.

Returning to our derivation of the MPS representation we reshape and suppress redundant rows and columns in the singular matrix and separate out the indices $(j_1, s_2)$ and $(s_2, j_2, j_3 \cdots, j_N)$:
\begin{equation}\label{eq:svd}
    c_{j_1, j_2, \cdots, j_N } = \sum_{s_2} {U_{j_1, s_2} S_{s_2,s_2} V^\dagger_{s_2, j_2, j_3, \cdots, j_N}}.
\end{equation}
We can now incorporate $S$ into $U_{j,s} \to A^{[j]}_{s}$ and $V^\dagger \to \widetilde{V}^\dagger$ as is convenient, where we have relabeled $U$ in accordance with notational convention:
\begin{equation}
    c_{j_1, j_2, \cdots, j_N } = \sum_{s_2} A^{[j_1]}_{s_2} \widetilde{V}^\dagger_{s_2, j_2, j_3, \cdots, j_N}.
\end{equation}
Repeating this procedure on $\widetilde{V}^\dagger$ across the next physical bi-partition using the combined indices $(s_2, j_2)$ and $(j_3, j_4, \cdots, j_N)$ yields
\begin{equation}
    c_{j_1, j_2, \cdots, j_N } = \sum_{s_2, s_3} A^{[j_1]}_{s_2} A^{[j_2]}_{s_2, s_3} \widetilde{V}^\dagger_{s_3, j_3, j_4, \cdots, j_N}.
\end{equation}
By continually decomposing the resulting $\widetilde{V}^\dagger$ we finally arrive at the MPS representation of our tensor $c$
\begin{equation}\label{eq:tensor-decomp}
    c_{j_1, j_2, \cdots, j_N } = \sum_{\{s\}} A^{[j_1]}_{s_1, s_2} A^{[j_2]}_{s_2, s_3} A^{[j_3]}_{s_3, s_4} \cdots A^{[j_N]}_{s_N, s_1}.
\end{equation}
The $s$ indices are `auxiliary' indices which connect neighbouring tensors and describe the internal degrees of freedom (thus they can be conveniently gauged). The auxiliary index $s_1$ connecting the first and final tensors has been inserted to account for closed boundary conditions; in the case of open boundary conditions it can be safely suppressed such that the first and final $A$ matrices become vectors.

The final state, by \cref{eq:initial} and \cref{eq:tensor-decomp}, is thus
\begin{equation}\label{eq:mps-decomp}
    |\text{MPS}[\psi, \chi]\rangle = \sum_{\{j\}, \{s\}} A^{[j_1]}_{s_1, s_2}A^{[j_2]}_{s_2, s_3}\cdots A^{[j_N]}_{s_{N}, s_1} |j_1, j_2, \cdots, j_n\rangle
\end{equation}
where the size of the $A$ matrices is limited by the bond dimension $\chi$ which in turn controls the fidelity of the MPS decomposition $|\langle\psi|\text{MPS}[\psi, \chi]\rangle|$. For clarity we must forego the usual notation $|\psi[A]\rangle$ for the MPS representation of a state as a parametrization in terms of the $A$ matrices, instead introducing new notation $|\text{MPS}[\psi, \chi]\rangle$ which clearly displays the bond dimension $\chi$ (rather than leaving it implicitly defined as the dimension of the $A$ matrices) and reframes the decomposition of a state into its MPS representation as a compression procedure rather than an exact parametrization:
\begin{equation}\label{eq:mps-compression}
    |\psi\rangle \to |\text{MPS}\left[\psi, \chi\right]\rangle.
\end{equation}
In the case that (i) the original state $|\psi\rangle$ has a low amount of entanglement, or that (ii) the bond dimension $\chi$ of the MPS representation is sufficiently high, the compression of \cref{eq:mps-compression} is close to lossless and the final MPS $|\text{MPS}\left[\psi, \chi\right]\rangle$ is close to the initial state $|\psi\rangle$.

We conclude this section with four pertinent parting notes. Firstly, that the decomposition of \cref{eq:mps-compression} is optimal in the sense that it minimizes the distance between the initial state and its MPS representation \cite{Orus2014}. Secondly, that the decomposition is not unique, as can be seen by simply gauging the bonds e.g. $A^{[j_1]}A^{[j_2]} = (A^{[j_1]}X)( X^{-1}A^{[j_2]}) = \widetilde{A}^{[j_1]}\widetilde{A}^{[j_2]} $; however this corresponds to a local change of basis and does not affect the physical properties of the MPS. Thirdly, that the number (and value) of the singular values across a bi-partition does not change under local operations, rendering it a genuine entanglement measure \cite{Sperling2011}. Finally, that the manifold of MPS states of a fixed bond dimension $\chi$ contains the manifolds of all MPS states of strictly lower bond dimension: with the $\chi=\Omega$ manifold being identical to the full Hilbert space and the $\chi=1$ manifold being identical to the set of all fully-separable states (no entanglement is present across any physical cut). The restructuring of Hilbert space into these nested manifolds, and the decomposition process of \cref{eq:mps-compression} on a generic state, are shown schematically in panel \textbf{b)} of \cref{fig:schematic}.


\section{Generalized Geometric Entanglement} 
\label{sec:generalizing}

The geometric entanglement $\mathcal{E}$ over pure states is defined as the distance between some pure state $|\psi\rangle$ and the nearest fully separable state $|\phi\rangle$ such that
\begin{equation} \label{eq:geo-ent}
    \mathcal{E} = \min_{\phi} \left[ 1-|\langle\psi|\phi\rangle|^2\right]
\end{equation}
is minimized over $|\phi\rangle \in \mathcal{S}$ where $\mathcal{S}$ is the space of all fully separable states \cite{Wei2003, DeChiara2018}. The prevailing generalization of this quantity considers instead minimization over $|\phi\rangle \in \mathcal{S}_k$ where $\mathcal{S}_k$ the set of all $k$-separable states $\mathcal{S}_k$. A $k$-separable state is a state that can be written as a product state of $k$ subsystems which may be internally entangled but do not share entanglement between them \cite{Blasone2008, Andreas2010, SenDe2010ggm}.
These quantities have seen widespread success, notably in the identification and analysis of bi-partite and genuine multi-partite entanglement \cite{Wei2003,Blasone2008,Shi2010, SenDe2010, Biswas2014,Sadhukhan2017, Singha2019}. Despite this, \cref{eq:geo-ent} and its immediate generalization in terms of $k$-separability have one major shortcoming: they cannot readily differentiate between simple and complicated entanglement structures. A product state of entangled Bell pairs, for example, will saturate \cref{eq:geo-ent} despite its trivial structure, and one can conceive of states with which are \textit{entirely} non-separable but have simple entanglement structures - e.g. the AKLT state \cite{Affleck1987, Schollwock2011}. Thus, whilst a generalization of the geometric entanglement from the perspective of separability is invaluable, there are contexts where a different generalization is more appropriate.

Our alternative generalization is simply the minimization of \cref{eq:geo-ent} over the manifold of MPS of fixed bond dimension $\chi$ insead of ($k$-)separable states:
\begin{equation}\label{eq:mps-geo}
   \mathcal{E}_\chi = 1 - |\langle\psi|\text{MPS}[\psi, \chi]\rangle|^2
\end{equation}
and is shown in diagrammatic tensor notation in panel \textbf{a)} of \cref{fig:schematic}. An overview of this notation is given in \cref{sec:app-diagram}. We have omitted the minimization from our notation because, as discussed in \cref{sec:mps} and noted in Ref.~\cite{Orus2014}, the minimization happens implicitly during the decomposition of \cref{eq:mps-compression}. Finally, given the discussion of the role of the bond dimension $\chi$ in \cref{sec:mps}, our generalization \cref{eq:mps-geo} is interpretable as the geometric entanglement from the perspective of entanglement complexity as opposed to $k$-separability. 

Intuitively, rather than organizing the full Hilbert space into nested sets of $k$-separable states like the existing generalizations of the geometric entanglement, the MPS formalism organizes it into nested manifolds of states with exact fixed-$\chi$ MPS representations (see panel \textbf{b)} of \cref{fig:schematic}). This picture moves us away from separability and towards the alternative, nuanced understanding of entanglement complexity given by the MPS formalism. This nested structure also implies, as every MPS manifold contains the manifolds of strictly lower bond dimension within it, the hierarchy $\mathcal{E}_1 \geq \mathcal{E}_2 \geq \cdots \mathcal{E}_\Omega$ where $\Omega$ is the total dimension of the full Hilbert space. Definitionally, and conveniently, the geometric entanglement of \cref{eq:geo-ent} and our generalization of \cref{eq:mps-geo} coincide $\mathcal{E}=\mathcal{E}_1$ at $\chi=1$ which defines a manifold of product states \cite{Schollwock2011}; this has been noted in Ref.~\cite{Teng2017} which uses the $\chi=1$ MPS representation to efficiently calculate the geometric entanglement - though it lacks the extension to higher bond dimensions. The restructuring of Hilbert space and interpretation of entanglement from the perspective of complexity rather than separability results in a quantity which, when it is extended to higher bond dimensions $\chi > 1$, captures behaviour which the geometric entanglement cannot.

Finally we remark that whilst MPS have been used in conjunction with the geometric entanglement before, these works focus on efficient calculation of existing measures, rather than in the construction of new measures. See e.g. Refs.~\cite{Shi2010, Teng2017, Singha2019}.

\section{The AKLT Model}
\label{sec:aklt}
In this section we demonstrate a situation in which the notion of separability is irrelevant: the detection of the non-separable AKLT ground state. Consider the spin-1 extended Haldane chain
\begin{equation}\label{eq:aklt-ham}
    H = \sum_j^n \vec{S}_j \cdot \vec{S}_{j+1} + \frac{J_\text{AKLT}}{3} \sum_j^n  \left(\vec{S}_j \cdot \vec{S}_{j+1}\right)^2
\end{equation}
where $\vec{S}_j = (S_j^x, S_j^y, S_j^z)^\top$ are vectors of spin-1 operators. At the point $J_\text{AKLT} = 1$, \cref{eq:aklt-ham} becomes the AKLT hamiltonian \cite{Affleck1987}, the ground state of which is a fused valence bond solid which is non-separable but admits an exact representation of an MPS of bond dimension $\chi=2$ \cite{Schollwock2011}:
\begin{equation}
    A^{[+]} = \sqrt{\frac{2}{3}}\sigma^+, \quad A^{[0]} = \frac{1}{\sqrt{3}}\sigma^z, \quad A^{[-]} = \sqrt{\frac{2}{3}}\sigma^-
\end{equation}
where the $\sigma^\pm$ and $\sigma^z$ operators are the standard pauli ladder and $z$ operators \cite{PerezGarcia2006, Orus2014}. In this setting it is clear that $\mathcal{E}_\chi$ should be able to detect the AKLT ground state, whilst the geometric entanglement and its $k$-separable generalizations should not.
\begin{figure}[ht]
    \centering
    \includegraphics[width=\linewidth]{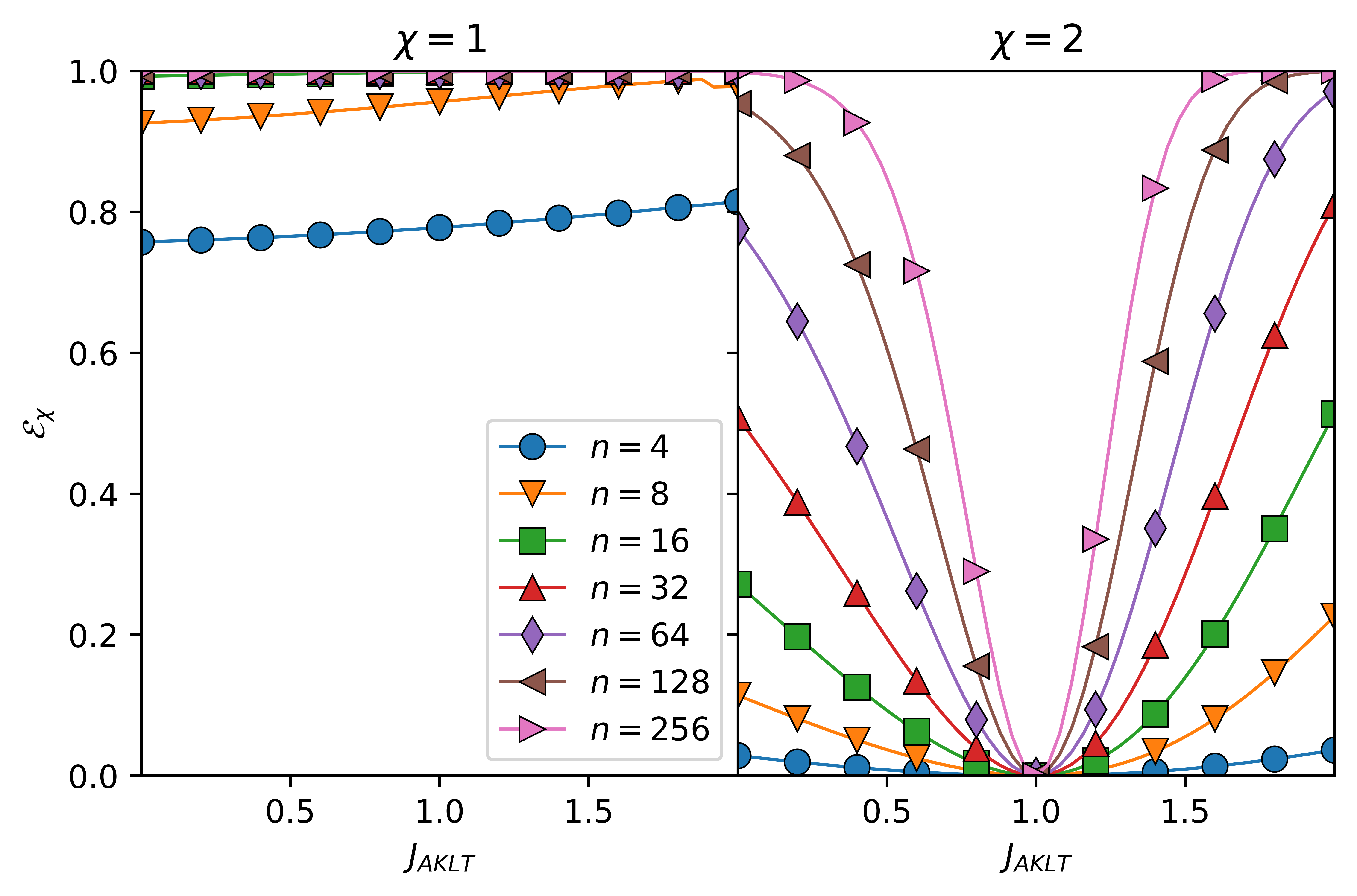}
    \caption{The \textbf{(left)} geometric entanglement $\mathcal{E}_1$ and \textbf{(right)} our generalized $\chi=2$ counterpart $E_2$ of the ground state of \cref{eq:aklt-ham} across the AKLT point. Only our generalization successfully locates the AKLT ground state.}
    \label{fig:aklt-results}
\end{figure}
We take open boundary conditions, and fix the elements of $\vec{S}_1$ and $\vec{S}_n$ (where $n$ is the system size) as spin-1/2 operators to lift the fourfold ground state degeneracy \cite{Wierschem2014}. We then probe the system's ground state using the geometric entanglement $\mathcal{E}_1$ and our first non-trivial generalization $\mathcal{E}_2$. Using DMRG implemented using a modification of quimb, we access large system sizes up to $n = 256$ \cite{White1993, Schollwock2005, Schollwock2011, Gray2018quimb}. The results are shown in \cref{fig:aklt-results}, from which we can see that the geometric entanglement $\mathcal{E}_1$ fails to detect the AKLT point at all, even in small systems in which it hasn't yet saturated to unity, whilst our $\chi=2$ generalization $\mathcal{E}_2$ successfully identifies the ground state in the thermodynamic limit. Additional results  of similar behaviour for the $J_1-J_2$ Antiferromagnetic isotropic Heisenberg model at the Majumdar-Ghosh ground point are given in \cref{sec:app-mg}.

\section{Ground State Phase Diagram of the Anisotropic Haldane Model}
\label{sec:haldane}
The fact that the toy problem of the previous section is best captured by $\mathcal{E}_2$ instead of $\mathcal{E}_1$ is - whilst an excellent demonstration of why our generalization is valuable - fairly obvious given the properties of the AKLT ground state. We now consider a more complicated situation in which higher generalizations $\chi \geq 2$ gradually reveal more and more details about the known phase diagram of a given system. This demonstrates the value of $\mathcal{E}_\chi$ as an exploratory tool for investigating systems which are not so thoroughly understood, e.g. in systems where optimal values of $\chi$ are not known \textit{a priori}. We consider an anisotropic Haldane chain
\begin{equation}\label{eq:haldane}
    H = J \sum_{j=0}^{L-1} \vec{S}_j \cdot \vec{S}_{j+1} + D \sum_{j=0}^L \left(S_j^z\right)^2 + E \sum_{j=0}^L  \left(S_j^x\right)^2 - \left(S_j^y\right)^2
\end{equation}
where the parameter $D$ tunes the strength of uniaxial anisotropies, and $E$ tunes the strength of rhombic anisotropies. The Hamiltonian of \cref{eq:haldane} is widely used, albeit often with one of the anisotropic terms set to zero, in the modelling of realistic spin systems \cite{DenNijs1989, Shaolong1992} (also see Ref.~\cite{Ren2018} and references therein).

\begin{figure}[ht]
    \centering
    \includegraphics[width=\linewidth]{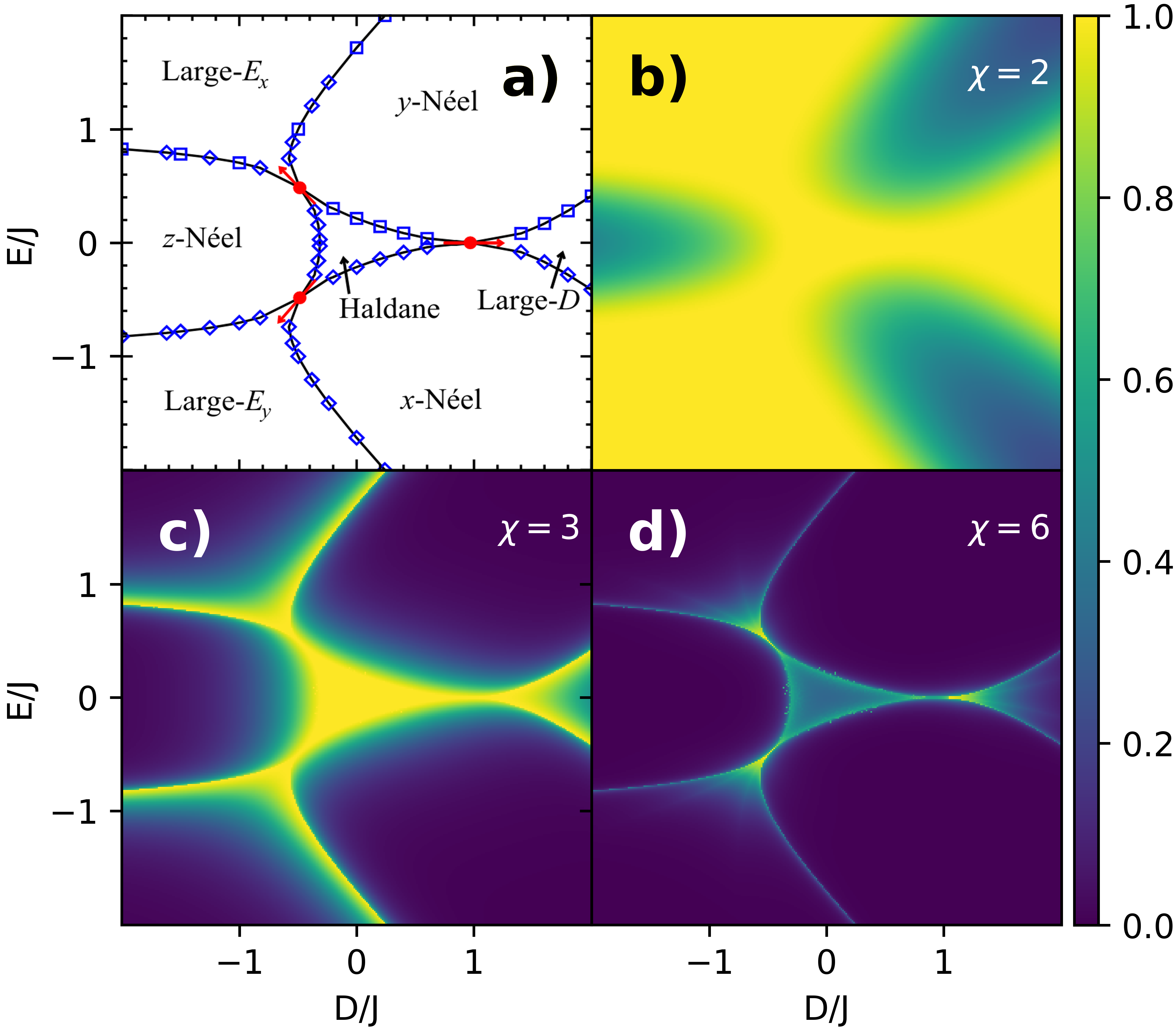}
    \caption{ The ground state phase diagram of the anisotropic Haldane model of \cref{eq:haldane}. Panel \textbf{a)} shows the phase diagram as determined in Ref.~\cite{Tzeng2017} (reproduced with permission). Panels \textbf{b)-d)} show $\mathcal{E}_\chi$ of the ground state for $\chi=2,3,6$ respectively. The ground states were calculated for a system of $n=512$ sites using two-site DMRG implemented using an extension of quimb \cite{White1993, Schollwock2005, Schollwock2011, Gray2018quimb}.}
    \label{fig:haldane-phases}
\end{figure}
The ground state phase diagram of the system as determined in Ref.~\cite{Tzeng2017} is shown in panel \textbf{a)} of \cref{fig:haldane-phases}. There are seven distinct gapless and gapful phases: the three N\'eel-like phases, the large-$E_x/E_y/D$ phases, and - most notably - the central gapped Haldane phase. We discuss these where relevant in the rest of this section. There are a litany of associated phase transitions in different universality classes, but we will only briefly mention the three Gaussian transitions between the Haldane phase and the large-$E_x/E_y/D$ phases (marked as red dots with arrows through them in panel \textbf{a)} of \cref{fig:haldane-phases}) \footnote{For a more detailed discussion of the phase transitions see Ref.~\cite{Tzeng2017} and the references therein.}. We consider an anti-ferromagnetic $J > 0$ coupling and so the ground state prefers maximal values $S_j^x = \pm 1$ everywhere, the different phases occur when this antiferromagnetic coupling and the anisotropies $D$ and $E$ assist or frustrate each other. The system is symmetric around $E=0$ as a negative $E$ simply corresponds to an inversion of $x$ and $y$ axes on each site. The ground states of each phase are best understood in terms of single-site ground states everywhere except the Haldane phase around $D=E=0$, and it is from this perspective that we discuss them below. The point $D=E=0$ itself is adiabatically connected to the AKLT ground state, as evidenced by the continuity of \cref{fig:aklt-results} across the interval $J_\text{AKLT} \in [0, 1]$, and thus the Haldane phase is best understood as having similar properties to the fused valence bond solid of the AKLT point.


We found (shown in panel \textbf{a)} of \cref{fig:app-haldane-phases} in \cref{sec:app-haldane}) that the geometric entanglement $\mathcal{E}_1$ is close to saturation across the entire region of phase space we investigate $D \in [0,2]$ and $E \in [0,2]$; this is simply due to the fact that - even in regions where single-site terms begin to dominate - the antiferromagnetic coupling still generates some entanglement. As the single-site terms dominate fully $D/J \to \infty$ or $E/J \to \infty$, the geometric entanglement once again becomes a useful investigative tool as the ground states become products of single-site ground states.

In contrast to the geometric entanglement, our generalizations $\mathcal{E}_\chi$ reveal more and more features of the phase diagram as a function of increasing $\chi$; a feature related to the fact that the different phases have different entanglement structures which are best captured by MPS of different bond dimensions. This is shown in panels \textbf{b)-d)} of \cref{fig:haldane-phases} where the phases of $\cref{eq:haldane}$ are captured by $\mathcal{E}_\chi$ for $\chi = 2,~3,~6$ respectively.

Panel \textbf{b)} shows the first non-trivial generalization $\mathcal{E}_2$ which successfully identifies states deep in the $x/y/z$-N\'eel phases. These phases occur when the system's ground state is close to a N\'eel state ($\chi=1$) of eigenstates $|S_j^{x/y/z} = \pm 1\rangle$ respectively. In the $z$-N\'eel phase this is assisted by low $E$ which prefers $|S_j^x = 0\rangle$ eigenstates and the antiferromagnetic coupling $J$ itself. In the $x$-N\'eel and $y$-N\'eel phases this is assisted by positive $D$ which prefers $|S_j^z = 0\rangle$. As such all three N\'eel states aren't frustrated away from their respective phase boundaries and these regions are revealed by low bond dimension $\chi=2$.

Panel \textbf{c)} shows $\mathcal{E}_3$ which reveals the full extent of the N\'eel phases and successfully identifies the large-$E_x/E_y/D$ phases. The former is due to slight frustration that each of the N\'eel phases experience close to their phase boundaries, an MPS of bond dimension $\chi=2$ simply doesn't capture enough information near these boundaries. The latter is due to the fact that each of the large-$E_x/E_y/D$ phases is frustrated. The large-$E_x/E_y$ phases have ground states close to product states of $|S_j^{x/y} = \pm 1\rangle$ but this is frustrated directly by negative $D$ and the antiferromagnetic coupling $J$ which prefer eigenstates $|S_j^z=\pm1\rangle$. The large-$D$ phase experiences a similar frustration, but entirely between the antiferromagnetic coupling and large positive $D$. We can also infer the existence of the Haldane phase around $D=E=0$, but not any of its properties or its phase boundaries.

Panel \textbf{d)} shows $\mathcal{E}_6$ which further narrows the phase boundaries and finally reveals the Haldane phase itself. A clear decrease of in $\mathcal{E}_6$ can be seen in the Haldane phase indicating that it is area-law entangled; a feature of the fact that the ground state at $D=E=0$ is adiabatically connected to the area-law AKLT ground state. In fact the AKLT ground state is a good approximation of the true ground state near $D=E=0$ in general \cite{CamposVenuti2006, Maximova2021}. The reason the Haldane phase is only captured by a slightly higher bond dimension $\chi=6$ compared to the other phases is simply due to the fact that all the terms of the Hamiltonian are of the same order, the system is thus highly frustrated, and is slightly more entangled - though it is still a valence bond solid similar in structure to the AKLT ground state.

Investigation of higher values of $\chi$ (shown in \cref{fig:app-haldane-phases} in \cref{sec:app-haldane}) reveals no new features aside from persistently high values of $\mathcal{E}_\chi$, only disappearing at $\chi=32$, at the Gaussian fixed points shown in panel \textbf{a)} of \cref{fig:haldane-phases} as red dots with arrows through them \cite{Chen2003}. For a more detailed discussion see \cref{sec:app-haldane}.

\begin{figure*}[ht]
        \centering
        \includegraphics[width=\textwidth]{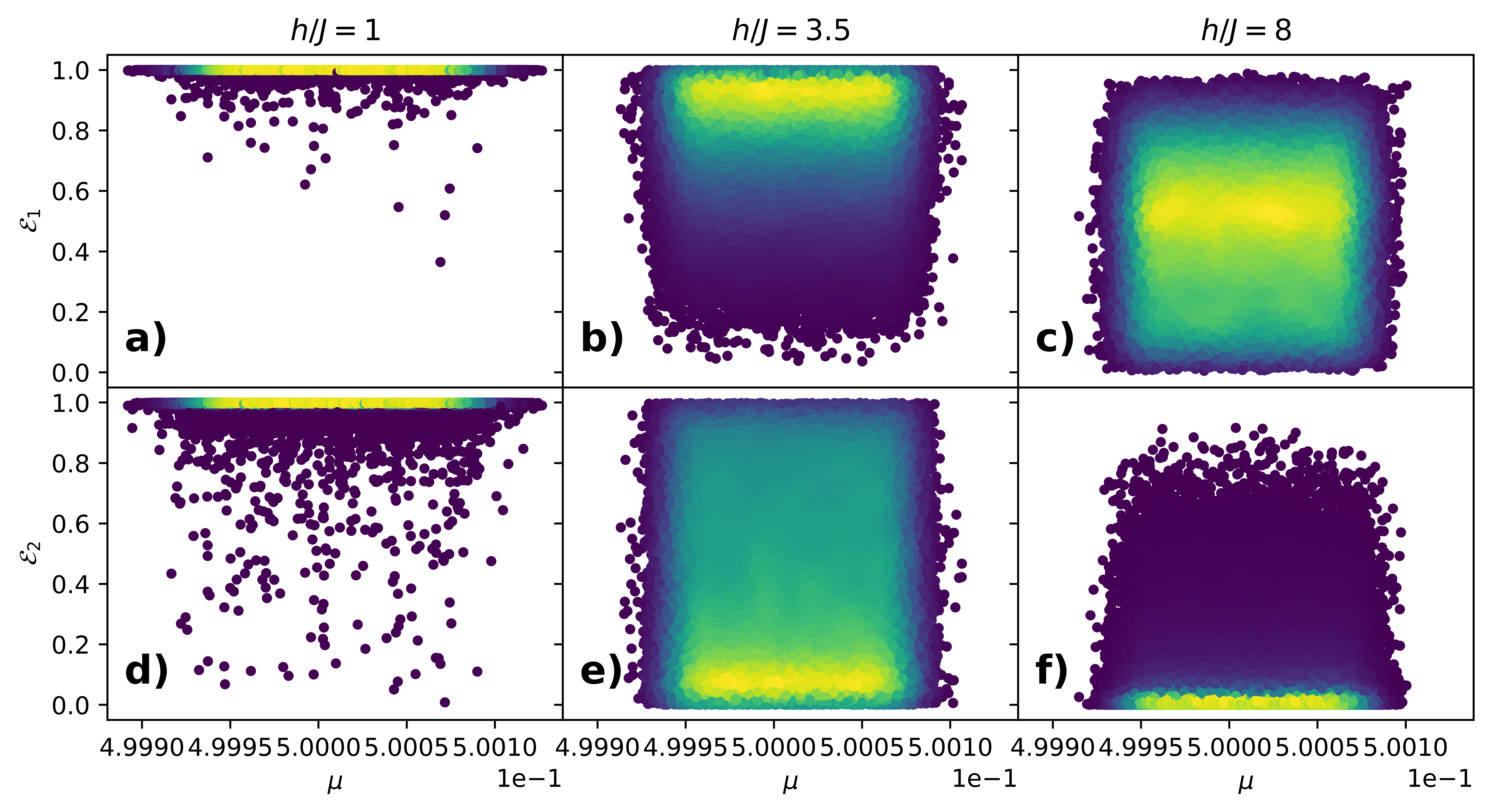}
        \caption{Each panel shows $\mathcal{E}_\chi$ of $100$ mid-spectrum eigenstates of $1024$ random realizations of \cref{eq:mbl-ham}. Each column shows a different disorder strength $h/J$ either \textbf{a),d)} in the ergodic regime $h/J=1$, \textbf{b),e)} the middle of the ergodic-MBL transition $h/J=3.5$, and \textbf{c),f)} in the MBL regime $h/J=8$. Each row shows either \textbf{a),b),c)} the conventional geometric entanglement $\chi=1$ or \textbf{d),e),f)} the first non-trivial generalization $\chi=2$. Brighter yellow coloration indicates a higher density of states. The dimensionless quantity $\mu \in [0,1]$ is the energy of the eigenstate relative to the extremal eigenenergies (i.e. $\mu=0$ is the ground state energy, and $\mu=1/2$ is the middle of the spectrum). }
        \label{fig:mbl-results-dmrg-scatter2}
\end{figure*}
\begin{figure}[ht]
    \centering
    \includegraphics[width=\linewidth]{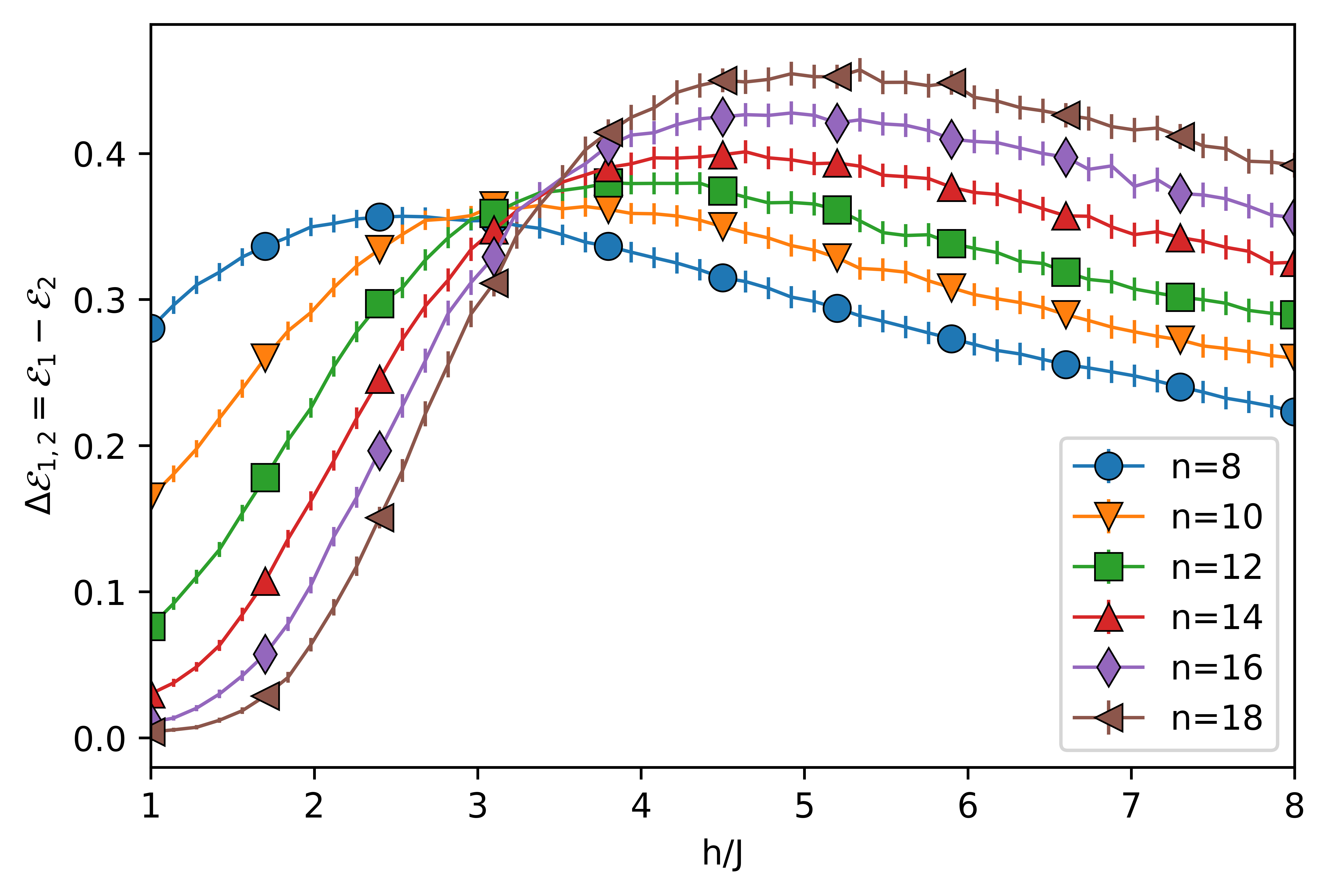}
    \caption{The difference $\Delta \mathcal{E}_{1,2} = \mathcal{E}_1 - \mathcal{E}_2$ between the conventional $\mathcal{E}_1$ and generalized $\mathcal{E}_2$ geometric entanglements across the ergodic-MBL transition. We can see clear peaks, as well as slow linear decay after the peaks with increasing $h/J$. The intersection of curves for the largest three systems near $h/J=3.5$ indicates scale-invariant behaviour near this point. }
    \label{fig:mblt_1_2}
\end{figure}

\section{Many Body Localization}
\label{sec:mbl}
Many body localization (MBL) is a mechanism by which an interacting many body quantum system fails to thermalize. The precise definition of thermalization in this context, though addressed in part by the eigenstate thermalization hypothesis \cite{Deutsch1991,Srednicki1994,Rigol2008}, is still debated (see Ref.~\cite{DAlessio2016} for a review); but certain hallmarks of MBL have been well established. Notable features of the MBL regime include: the breakdown of internal energy and particle transport, the emergence of local memory and local integrals of motion, and mid-spectrum eigenstates exhibiting area-law like entanglement entropy (see Refs.~\cite{Nandkishore2015, Alet2018, Abanin2017}). This final point is what we investigate here: area-law MBL states, whilst rich and generally non-separable, have efficient representations as low bond dimension MPS.
Thus our generalization $\mathcal{E}_\chi$ is a natural measure of bulk entanglement in MBL systems, and should readily detect MBL eigenstates. It is worth explicitlly noting here that the MBL transition takes place across the entire spectrum and is a departure from the ground state transitions we have considered thus far.

Consider the prototypical spin-$1/2$ Heisenberg hamiltonian with quenched z-field disorder
\begin{equation}\label{eq:mbl-ham}
H = J \sum_j \vec{S}_j \cdot \vec{S}_{j+1} + \sum_j h_j S_j^z
\end{equation}
where $\vec{S}_j$ are vectors of the standard spin-1/2 spin operators and the $h_j$ are quenched random fields box distributed in the interval $h_j \in [-h, h]$. The ratio of disorder strength to Heisenberg coupling $h/J$ tunes the model; for large disorder $h/J \gg 1$, the system is localized. We consider systems of size up to $n = 18$ which, whilst too small to extract reliable thermodynamic properties of MBL through e.g. conventional scaling analyses, allows us to differentiate ergodic and localized regimes \cite{Sierant2020, Panda2020}. 

We first investigate $\mathcal{E}_\chi$ for individual mid-spectrum eigenstates across the MBL transition using both the conventional geometric entanglement $\mathcal{E}_1$ in panels \textbf{a)-c)} and our first non-trivial generalization $\mathcal{E}_2$ in panels \textbf{d)-f)} of \cref{fig:mbl-results-dmrg-scatter2}. Each panel shows $\mathcal{E}_\chi$ for $1024$ samples of $100$ mid-spectrum eigenstates, for a total of $102400$ data points, these are then coloured according to a Gaussian kernel density estimation. From panels \textbf{a)} and \textbf{d)} we can see that both $\mathcal{E}_1$ and $\mathcal{E}_2$ are high in the ergodic regime $h/J=1$, implying the well-known property that mid-spectrum eigenstates of generic hamiltonians are volume-law and thus have no efficient representation as low bond dimension MPS. Panels \textbf{b)} and \textbf{e)} indicate that, close to the transition point $h/J = 3.5$, $\mathcal{E}_1$ remains high, but the average value of $\mathcal{E}_2$ - despite the existence of many individual eigenstates which have $\mathcal{E}_2$ far from zero - drops suddenly. This implies that eigenstates are far from product states, but are starting to become area-law entangled as low dimension MPS representations become increasingly viable. Finally panels \textbf{c)} and \textbf{f)} show slightly lower values of $\mathcal{E}_1$ and near-zero values of $\mathcal{E}_2$ in the MBL regime $h/J = 8$. This indicates that, in addition to almost all the eigenstates being entirely area-law entangled, many of the states also have considerable overlap with product states. This is evidence that we are witnessing the onset of behaviour similar to the $h/J \to \infty$ case where all ground states simply become product states of local $S^z_j$ eigenstates.

Given the results of \cref{fig:mbl-results-dmrg-scatter2} and the associated discussion, we notice that $\mathcal{E}_1$ and $\mathcal{E}_2$ coincide in the ergodic phase $\mathcal{E}_1=\mathcal{E}_2=1$, diverge near the ergodic-MBL transition point, and should coincide again deep in the MBL phase $\mathcal{E}_1=\mathcal{E}_2=0$. This is due to the fact that MPS of bond dimension $\chi=1$ and $\chi=2$ are both equally bad representations of thermal states on the ergodic side of the transition, and both equally exact representations of product states on the extreme $h/J \to \infty$ MBL side of the transition. This behaviour is captured by an equation of the form
\begin{equation}\label{eq:rel-geo}\Delta \mathcal{E}_{\chi_{1},\chi_{2}} = \mathcal{E}_{\chi_{1}}-\mathcal{E}_{\chi_{2}}.\end{equation}
which is strictly non-negative and bounded in the interval $[0, 1]$ for $\chi_1 < \chi_2$. Quantitatively \cref{eq:rel-geo} captures how much the fidelity of the MPS representation of a given state improves when we increase the bond dimension $\chi_1 \to \chi_2$ \footnote{For a related discussion see \cref{sec:app-mblt}.}.

We restrict ourselves to an analysis of the $\chi_1 = 1$ and $\chi_2 = 2$ case in the main body of this paper, but include an analysis of up to bond dimension $\chi_2 = 16$ in \cref{sec:app-mblt}. In all cases we average $\Delta \mathcal{E}_{1,2} = \mathcal{E}_1-\mathcal{E}_2 \geq 0$ over $512$ disorder samples (realizations of the Hamiltonian of \cref{eq:mbl-ham}) and $10$ mid-spectrum eigenstate samples for each disorder sample. The results of this analysis are shown in \cref{fig:mblt_1_2} where we can clearly see $\Delta \mathcal{E}_{1,2} = 0$ in the ergodic regime, and $\Delta \mathcal{E}_{1,2}$ decreasing linearly towards zero deep in the localized regime. In the transition region we can see a crossover point around $h/J \sim 3.5$ (considering the largest three sizes available) indicating scale-invariant behaviour around the region where the critical point $h_c \geq 3.5$ is usually found for similar small systems in the canonical model of \cref{eq:mbl-ham} \cite{Khemani2017}. We also note a slight drift of this crossover which is not an atypical pathology in extant analyses at similar scales. Whilst $\mathcal{E}_\chi$, and by extension \cref{eq:rel-geo} cannot diverge by definition, its gradient can: a feature we can see clearly in \cref{fig:mblt_1_2} close to $h/J=3.5$ with steeper gradients for larger system sizes. We found similar behaviour in $\Delta \mathcal{E}_{1,\chi_2}$ up to $\chi_2 \leq 16$, results for which are shown and in \cref{fig:app-mblt-4panel} in \cref{sec:app-mblt}; indicating that the coarse-graining of entanglement found in MPS representations of low bond dimension does not erase the qualitative features of the transition.

\section{Conclusions}
\label{sec:conclusions}

In this paper we have introduced a scalable quantification of geometric entanglement that doesn't appeal to separability. Rather, through the MPS formalism and the bond dimension $\chi$, this approach encourages an alternative understanding of entanglement in terms of entanglement complexity: the efficiency of state representation under entanglement coarse-graining. This change in perspective yields a generalized geometric measure of entanglement $\mathcal{E}_\chi$ which succeeds in contexts where the conventional geometric entanglement (coincident with $\mathcal{E}_1$) and its immediate $k$-separable generalization cannot. We additionally note that, due to the advantageous fact that $\mathcal{E}_\chi$ is still derived from an overlap between two states, it retains the positive feature of being an experimentally measurable quantity through a SWAP test \cite{Barenco1997,Buhrman2001}. We have demonstrated the value of $\mathcal{E}_\chi$ in a variety of different contexts. Firstly, at the AKLT point, which can be detected by $\mathcal{E}_2$ but not by $\mathcal{E}_1$. Secondly, in a more exploratory setting, we found that the phases of the anisotropic Haldane model, each having their distinct signature in  entanglement-complexity, are gradually revealed by $\mathcal{E}_\chi$ as we vary $\chi$. And finally in the context of MBL where, away from ground state analyses, $\mathcal{E}_\chi$ and relative entanglements $\Delta \mathcal{E}_{\chi_1,\chi_2}$ give us a tunable quantification of the transition between volume and area law entangled eigenstates across the spectrum. 

\section{Acknowledgements}

A. N.-K. thanks Y.-C. Tzeng for his valuable insights into the anisotropic Haldane model, A. Bayat and P. Bradshaw for their assistance on paper structure and content, A. Green and A. Pal for their early advice on MPS and the geometric entanglement, and F. Azad and L. Gover for their insight regarding the practical use of tensor networks. The authors acknowledge the EPSRC grant Nonergodic quantum manipulation EP/R029075/1.
\bibliography{refs}

\appendix


\section{Diagrammatic Tensor Notation}
\label{sec:app-diagram}

Diagrammatic tensor notation, alternatively named Penrose notation for its originator Roger Penrose in Ref.~\cite{Penrose1971}, is a visual depiction of tensors and tensor networks. The mathematical rules of tensor manipulation have corresponding diagrammatic representations and so the tedious process of calculation and tally-keeping of indices is abstracted away into doodles that are readable at a glance.

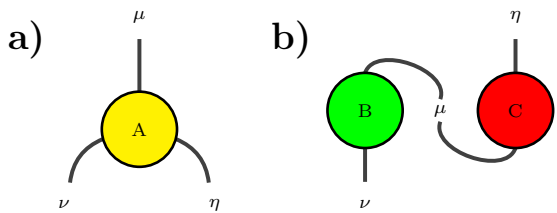
\begin{figure}[ht]
    \centering
    \begin{tikzpicture}
        
        \Vertex[x=0, y=0, color=yellow, size=1, label=A]{Atensor};
        \Vertex[x=0, y=1.5, Pseudo, label=$\mu$]{Amu};
        \Vertex[x=-1, y=-1, Pseudo, label=$\nu$]{Anu};
        \Vertex[x=1, y=-1, Pseudo, label=$\eta$]{Aeta};
        \Edge(Atensor)(Amu)
        \Edge[bend=-30](Atensor)(Anu) 
        \Edge[bend=30](Atensor)(Aeta)

        \Vertex[x=3, y=.25, color=green, size=1, label=B]{Btensor};
        \Vertex[x=5, y=.25, color=red, size=1, label=C]{Ctensor};
        \Vertex[x=4, y=.25, color=red, size=.3, Pseudo, label=$\mu$]{Mtensor};
        \Vertex[x=3, y=-1, Pseudo, label=$\nu$]{Bnu};
        \Vertex[x=5, y=1.5, Pseudo, label=$\eta$]{Ceta};
        
        \Edge[bend=90](Btensor.north)(Mtensor.north)
        \Edge[bend=-90](Mtensor.south)(Ctensor.south)
        \Edge(Btensor)(Bnu) 
        \Edge(Ctensor)(Ceta)
        \Text[x=-1., y=1.2, width = 1.5cm,fontsize=\Large]{\textbf{a)}}
        \Text[x=2.5, y=1.2, width = 1.5cm,fontsize=\Large]{\textbf{b)}}

    \end{tikzpicture}
    \caption{The basics of diagrammatic notation. Panel \textbf{a)} shows the diagrammatic representation of the tensor $A^{\mu}_{\nu\mu}$, and \textbf{b)} shows the diagrammatic representation of the contraction $B^\mu_\nu C_\mu^\eta$. }
    \label{fig:app-schematic-penrose}
\end{figure}

The basic object, the tensor itself, is represented as a blob with legs attached: each leg corresponds to an index of that tensor. Legs going up correspond to contravariant indices (up indices), and legs going down correspond to covariant indices (down indices); though this is often ignored for indices that are contracted over or in situations where the difference is irrelevant. The tensor $A^{\mu}_{\nu\eta}$ for example is shown in panel \textbf{a)} of \cref{fig:app-schematic-penrose}. Contractions over pairs of indices are drawn simply by connecting the indices in question; the tensor network $B^\mu_\nu C_\mu^\eta$ for example is shown in panel \textbf{b)} of \cref{fig:app-schematic-penrose}. 

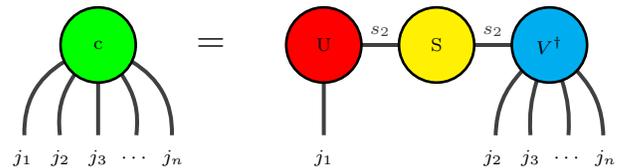
\begin{figure}[ht]
    \centering
    \begin{tikzpicture}
        
        \Vertex[x=0, y=.5, color=green, size=1, label=c]{ctensor};
        \Vertex[x=1.5, y=.5, Pseudo, fontsize=\Large, label={$=$}]{equals};
        \Vertex[x=3, y=.5, color=red, size=1, label={U}]{U};
        \Vertex[x=4.5, y=.5, color=yellow, size=1, label={S}]{S};
        \Vertex[x=6, y=.5, color=cyan, size=1, label={$V^\dagger$}]{Vd};
        
        \Vertex[x=-1,y=-1, Pseudo, label=$j_1$]{j1left};
        \Vertex[x=-.5,y=-1, Pseudo, label=$j_2$]{j2left};
        \Vertex[x=0,y=-1, Pseudo, label=$j_3$]{j3left};
        \Vertex[x=.5,y=-1, Pseudo, label=$\cdots$]{dotsleft};
        \Vertex[x=1,y=-1, Pseudo, label=$j_n$]{jnleft};

        \Vertex[x=3,y=-1, Pseudo, label=$j_1$]{j1right};
        \Vertex[x=5.25,y=-1, Pseudo, label=$j_2$]{j2right};
        \Vertex[x=5.75,y=-1, Pseudo, label=$j_3$]{j3right};
        \Vertex[x=6.25,y=-1, Pseudo, label=$\cdots$]{dotsright};
        \Vertex[x=6.75,y=-1, Pseudo, label=$j_n$]{jnright};
        
        \Edge[bend=30](j1left)(ctensor);
        \Edge[bend=20](j2left)(ctensor);
        \Edge(j3left)(ctensor);
        \Edge[bend=-20](dotsleft)(ctensor);
        \Edge[bend=-30](jnleft)(ctensor);

        \Edge(j1right)(U);
        \Edge[bend=20](j2right)(Vd);
        \Edge[bend=10](j3right)(Vd);
        \Edge[bend=-10](dotsright)(Vd);
        \Edge[bend=-20](jnright)(Vd);
        
        \Edge[label=$s_2$,position=above](S)(Vd);
        \Edge[label=$s_2$,position=above](U)(S);

    \end{tikzpicture}
    \caption{A diagrammatic version of the singular value decomposition of \cref{eq:svd}.}
    \label{fig:app-schematic-svd}
\end{figure}

The standard tensor manipulations then become simple diagrammatic tricks. For example, raising or lowering an index via the metric tensor corresponds to extending the corresponding leg until it points upwards or downwards. More complicated algorithms can also be readily represented, consider the singular value decomposition of \cref{eq:svd} (in which we have omitted any notion of upper or lower indices), which has been drawn diagrammatically in \cref{fig:app-schematic-svd}. Note we have the contraction over the single index $s_2$ is represented by two contractions over the same index $s_2$ going into and out of $S$, this can be understood either as equivalent to the promotion of $S$ to a diagonal tensor of increased rank, or as notational convenience in representing a contraction over three indices.

\section{The Majumdar-Ghosh Point}
\label{sec:app-mg}
Consider the $J_1 - J_2$ model defined by the Hamiltonian
\begin{equation}\label{eq:mg-ham}
    H = J_1 \sum_j^n \vec{S}_j \cdot \vec{S}_{j+1} + J_2 \sum_j^n  \vec{S}_j \cdot \vec{S}_{j+2}
\end{equation}
where $\vec{S}_j$ are vectors of standard spin-1/2 operators, across the Majumdar-Ghosh point $J_2 = J_1/2$. At this point, for open boundary conditions, the bulk of the unique ground state is a simple valence bond solid: a product state of singlets \cite{Majumdar1969}. Such a ground state has an exact representation as an MPS of bond dimension $\chi=2$ and is $n/2$-separable.

\begin{figure}[ht]
    \centering
    \includegraphics[width=\linewidth]{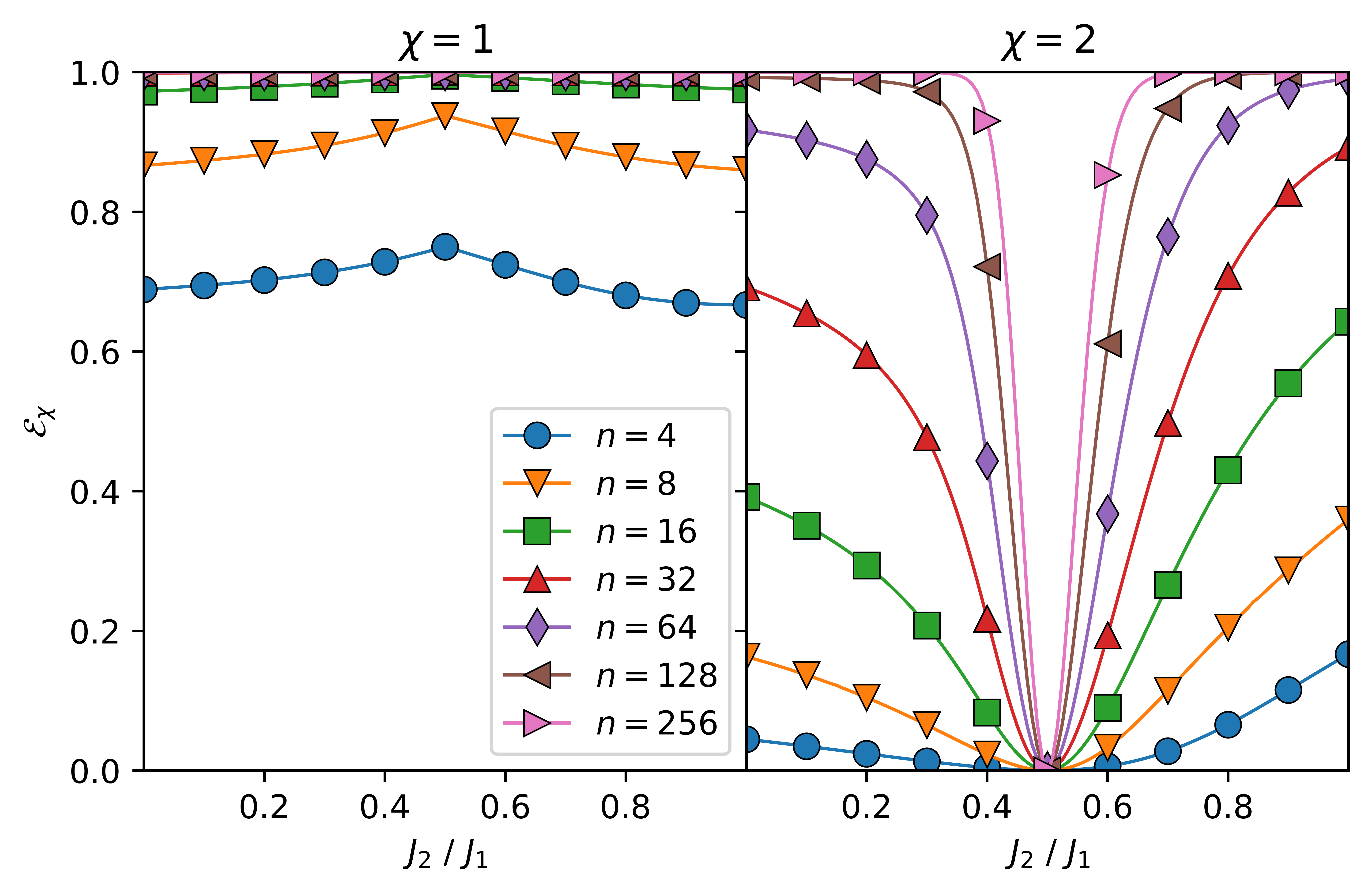}
    \caption{The \textbf{(left)} geometric entanglement $\mathcal{E}_1$ and \textbf{(right)} its first non-trivial generalization $\mathcal{E}_2$ of the $J_1-J_2$ ground state across the Majumdar-Ghosh point. Whilst the conventional geometric entanglement initially shows a small peak at the Majumdar-Ghosh point, only $\mathcal{E}_2$ successfully locates the ground state in the thermodynamic limit.}
    \label{fig:app-mg-results}
\end{figure}

We fix $J_1=1$ and investigate the generalized geometric entanglements $\mathcal{E}_1$ and $\mathcal{E}_2$ across the Majumdar-Ghosh point using two-site DMRG implemented using quimb to reach ground states of \cref{eq:mg-ham} for large system sizes ($n = 256$). These results are shown in \cref{fig:app-mg-results}, from which we can see clearly that - despite an initial peak at small system sizes - the conventional geometric entanglement completely fails to identify the point in the thermodynamic limit, whilst the $\chi=2$ generalization successfully captures the expected behaviour. Despite the clear advantage of $\mathcal{E}_2$ in this context, existing generalizations of the geometric entanglement based on $k$-separability can also detect the $n/2$-separable Majumdar-Ghosh ground state.

\section{Additional Results for the Anisotropic Haldane Model}
\label{sec:app-haldane}

\begin{figure}[ht]
    \centering
    \includegraphics[width=\linewidth]{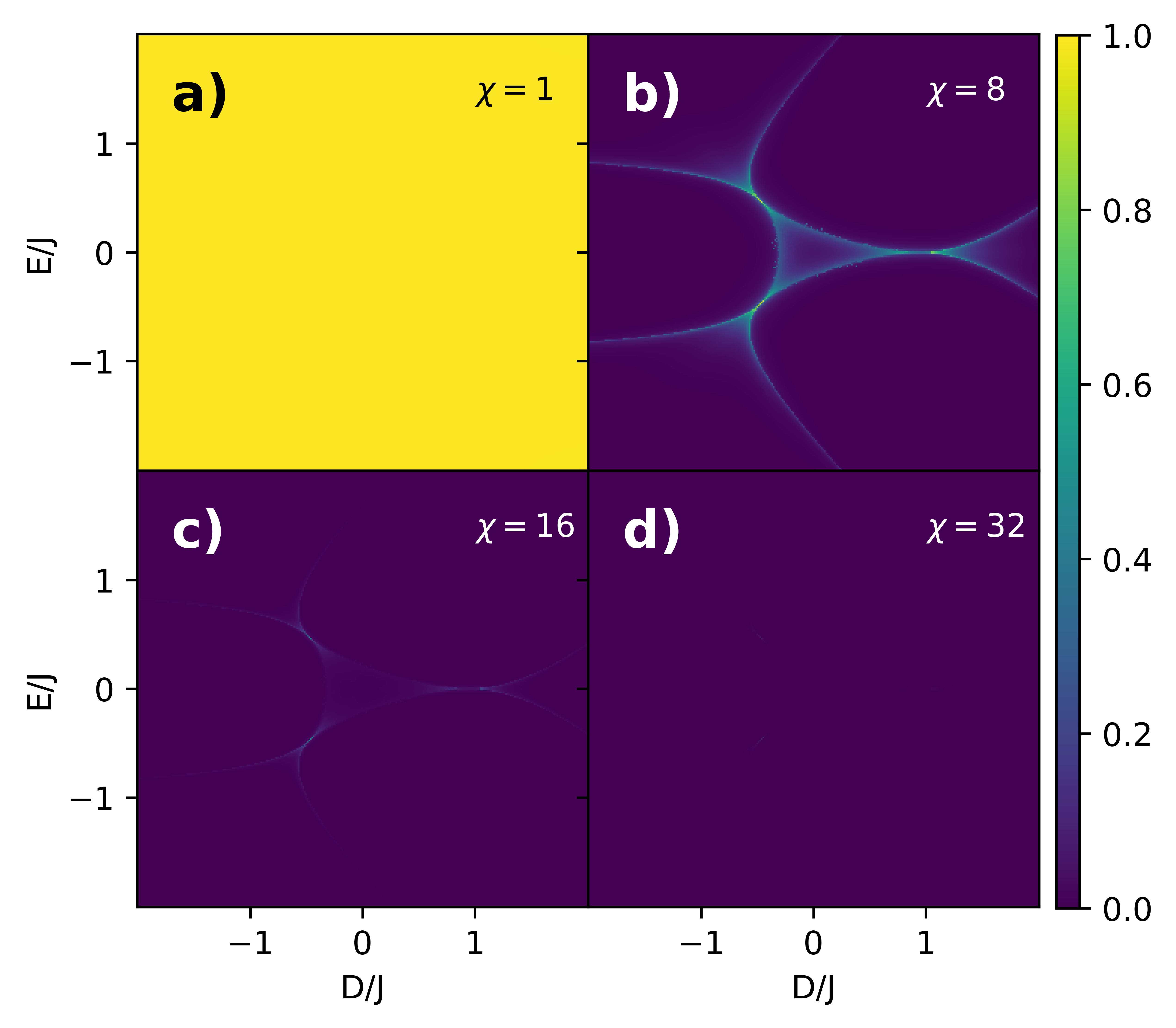}
    \caption{The ground state phase diagram of the anisotropic Haldane model of \cref{eq:haldane}. Panel \textbf{a)} shows the geometric entanglement $\mathcal{E}_1$. Panels \textbf{b)-d)} show $\mathcal{E}_\chi$ of the ground state for $\chi=8,16,32$ respectively. The ground state was calculated for a system of $n=512$ sites using two-site DMRG implemented using an extension of quimb \cite{Gray2018quimb}.}
    \label{fig:app-haldane-phases}
\end{figure}

This section concerns itself with the ground state phase diagram of the anisotropic Haldane model of \cref{eq:haldane} as discussed in \cref{sec:haldane} of the main text. In addition to the generalized geometric entanglement $\mathcal{E}_\chi$ of the ground states for $\chi=2,3,6$ considered in the main text and shown in panels \textbf{b)-d)} of \cref{fig:haldane-phases}, we provide results here for the geometric entanglement itself $\mathcal{E}_1$ and larger values of $\chi=8,16,32$. These results are shown in \cref{fig:app-haldane-phases}.

As noted in the main text, panel \textbf{a)} of shows that the geometric entanglement $\mathcal{E}_1$ is close to saturation over the entire region we investigate. This is due to the fact that the antiferromagnetic coupling $J$ generates some entanglement. Panel \textbf{b)} shows $\mathcal{E}_8$ in which the Haldane phase has become very clearly defined, reinforcing the idea that - whilst it is more entangled than the other phases' ground states and the AKLT state - it is still area-law entangled and admits a low-$\chi$ MPS representation as expected. Panel \textbf{c)} shows $\mathcal{E}_{16}$ in which only the critical regions near the Gaussian critical points from the Haldane phase to the large-$E_x/E_y/D$ phases are visible; this aligns with the understanding that - close to criticality - low-$\chi$ MPS representations generally fail. Panel \textbf{b)} indicates that an MPS of bond dimension $\chi=32$ is enough to represent even ground states near the critical regions; going to finer grid resolutions or larger system sizes may frustrate this however.

\section{Additional Analyses of the MBL Transition}
\label{sec:app-mblt}
In this section we discuss some additional results of the investigation of the ergodic-MBL transition found in \cref{sec:mbl} of the main text. We consider the relative generalized geometric entanglement $\Delta \mathcal{E}_{1, \chi} = \mathcal{E}_1-\mathcal{E}_\chi$ (defined in \cref{eq:rel-geo}) for $\chi \geq 2$, unlike the $\chi=2$ case considered in the main text. First we note that due to the restructuring of Hilbert space into a hierarchy of nested manifolds of MPS with fixed bond dimension there is an associated hierarchy $\mathcal{E}_1 \geq \mathcal{E}_2 \geq \cdots \geq \mathcal{E}_\Omega$ in the generalized geometric entanglement (where $\Omega$ is the dimension of the total Hilbert space). Two corollaries to this fact are: (i) that $\Delta \mathcal{E}_{1,\chi} \geq 0$ with equality only when the state in question is a product state or when $\chi=1$, and (ii) that there exists a similar hierarchy in $\Delta \mathcal{E}_{1,\chi}$:
\begin{equation}\label{eq:app-rel-hierarchy}
    \Delta \mathcal{E}_{1,2} \leq \Delta \mathcal{E}_{1,3} \leq \cdots \leq \Delta \mathcal{E}_{1,\Omega}.
\end{equation}
\begin{figure}[ht]
    \centering
    \includegraphics[width=\linewidth]{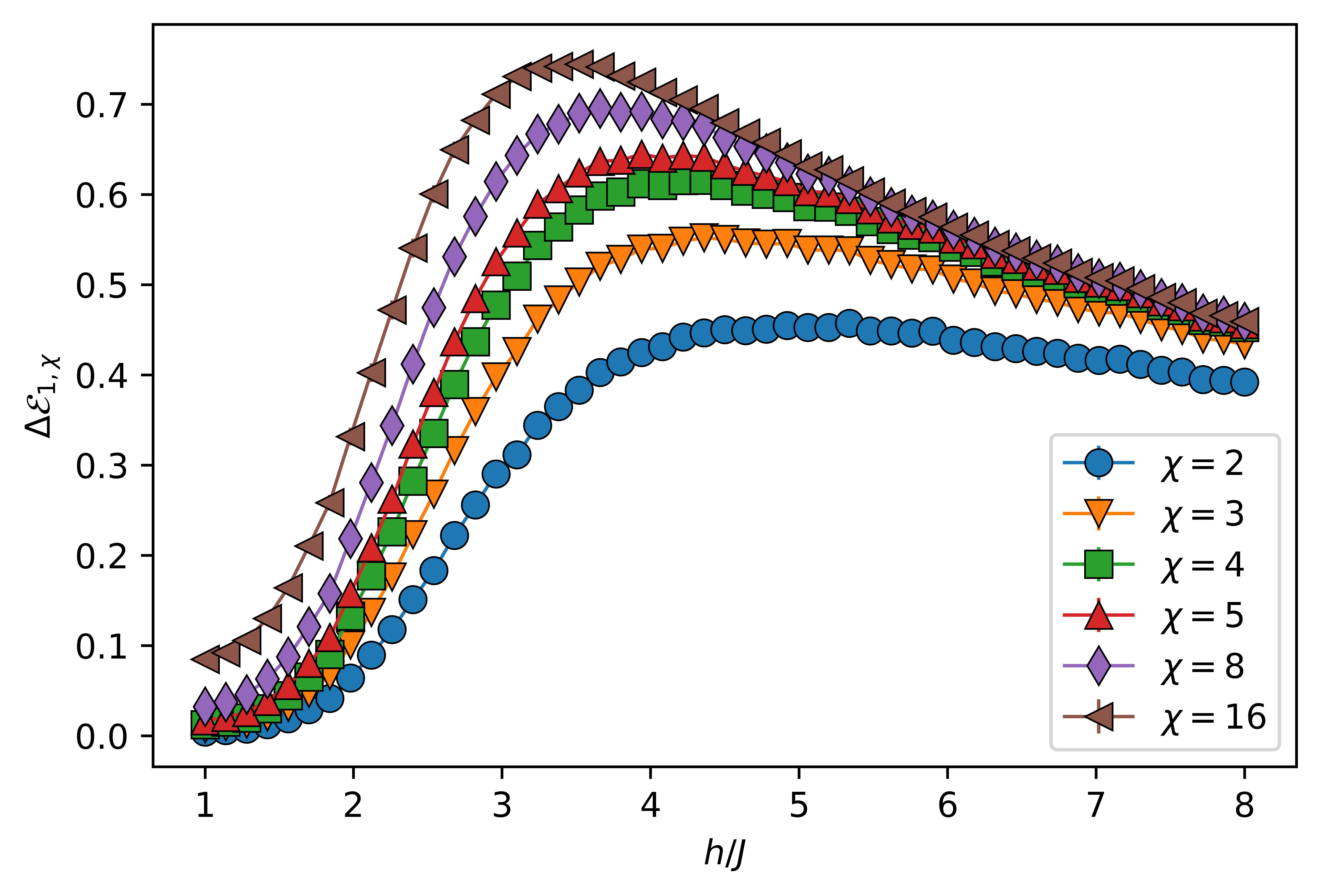}
    \caption{The relative generalized geometric entanglement $\Delta \mathcal{E}_{1,\chi}$ across the ergodic-MBL transition for different values of $\chi$ in a system of size $n=18$. We see clear peaks in the transition region close to $h/J=3.5$, as well as slow linear decay after the peaks with increasing $h/J$.}
    \label{fig:app-mblt-fixedn}
\end{figure}
In all cases we average $\Delta \mathcal{E}_{1,\chi}$ over $512$ disorder samples (realizations of the Hamiltonian of \cref{eq:mbl-ham}) and $10$ mid-spectrum eigenstate samples for each disorder sample; calculations carried out using quimb \cite{Gray2018quimb}. The hierarchy of \cref{eq:app-rel-hierarchy} is shown for a system of size $n=18$ in \cref{fig:app-mblt-fixedn}, and in addition we see a clear peak in the critical region near $h/J = 3.5$. This supports the argument made in the main text that low-$\chi$ MPS representations are equally bad in the ergodic regime as even $\chi=16$ MPS are poor, and become equally good in the MBL regime. 
\begin{figure}
    \centering
    \includegraphics[width=\linewidth]{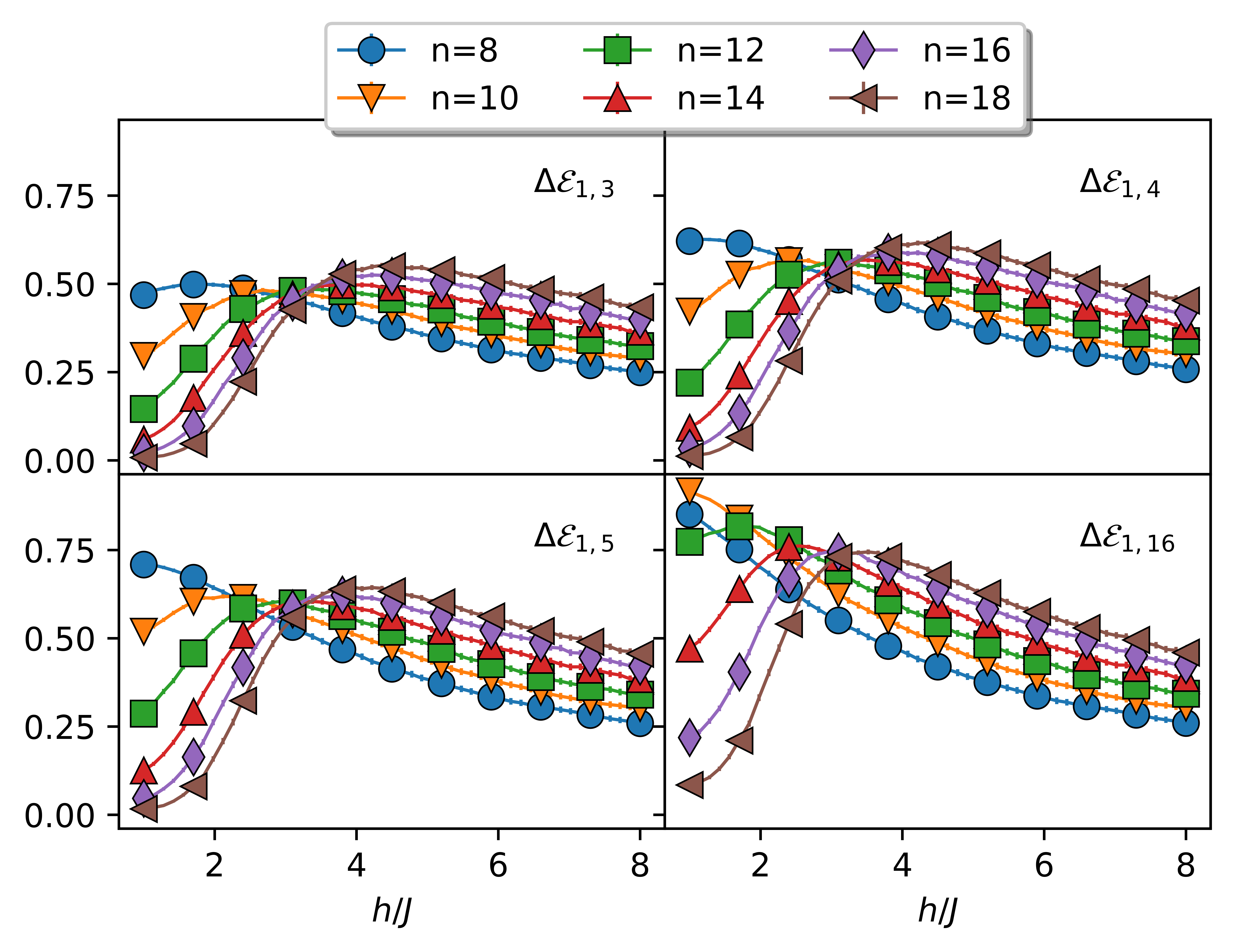}
    \caption{The relative generalized geometric entanglement $\Delta \mathcal{E}_{1,\chi}$ across the ergodic-MBL transition for different values of $\chi$ and different sizes. We see clear peaks and an intersection in the transition region near $h/J=3.5$, as well as slow linear decay after the peaks with increasing $h/J$. This indicates scale-invariant behaviour close to criticality.}
    \label{fig:app-mblt-4panel}
\end{figure}
We also investigate $\Delta \mathcal{E}_{1,\chi}$ for a range of different system sizes, the results of which are shown in \cref{fig:app-mblt-4panel}. From this figure we can see that the characteristic intersection of lines indicating scale-invariance close to $h/J=3.5$ and the drift of this point with increasing $n$ noted in the discussion of \cref{fig:mblt_1_2} in the main text are also present here. The results of this section motivated the exclusive use of $\Delta\mathcal{E}_{1,2}$ in the main text: no new qualitative information is revealed by accessing higher values of $\chi$ aside from a more pronounced critical peak in \cref{fig:app-mblt-fixedn}. Whether or not different $\Delta\mathcal{E}_{\chi_1,\chi_2}$ yield different quantitative results under a sophisticated scaling analysis or in other situations is beyond the scope of this paper.

\end{document}